**Putnam looks at quantum mechanics (again and again)**

Christian Wüthrich
University of California, San Diego

**Abstract**

Hilary Putnam (1965, 2005) has argued that from a realist perspective, quantum mechanics stands in need of an interpretation. Ironically, this hypothesis may appear vulnerable against arguments drawing on Putnam's own work. Nancy Cartwright (2005) has urged that his 1962 essay on the meaning of theoretical terms suggests that quantum mechanics needs no interpretation and thus stands in tension with his claim of three years later. She furthermore contends that this conflict should be resolved in favour of the earlier work, as quantum mechanics, like all successful theories, does not need an interpretation. The first part of this essay deflates both of these objections. The second part addresses and evaluates Putnam's own assessments of the main interpretative options available in 1965 and 2005. Although we may disagree on some aspects, his pessimistic conclusion will come out largely unscathed, and, in fact, enhanced. I will close by briefly stating the historical relevance of this work.

## 1 Introduction

Quantum mechanics offers the deepest and most compellingly confirmed theory science has ever devised. Yet, without a doubt, it also continues to surprise, puzzle, indeed stupefy us to an unparalleled extent. It frustrates our most dearly held intuitions concerning the material constitution of our world and seemingly demands their radical reconstitution. At the heart of the challenge quantum mechanics poses to our intuitive conceptions—even if tutored by classical physics, and indeed by relativity theory—we find the so-called 'measurement problem' and the non-local connections apparently implied by the experimental disconfirmation of Bell's inequality. The former concerns the interpretation of quantum mechanics itself, i.e., the question of what precisely the theory claims to be the case in our world; the latter strains the theory's relationship to another highly successful theory of physics, special relativity. Throughout his career, Hilary Putnam has reflected upon the foundations of quantum mechanics, particularly addressing the measurement problem and its ramifications.[1]

[1] Without making some valiant, but vain, attempt at being comprehensive here, the reader would be well advised to use his articles of 1965 and 2005 as useful

As Putnam has realized early on, the measurement problem in quantum mechanics arises particularly sharply for advocates of scientific realism, i.e., for those who believe that the stunning predictive success of quantum mechanics can only be explained if quantum mechanics is at least approximately true. This saddles the scientific realist with the task of explicating the mathematical formalism of the theory in terms of its reference to the actual world. For example, quantum mechanics represents the state of a physical system by so-called 'wave functions'. On a realist (or "nonoperationist") view, for Putnam, offering an answer to what the significance of these 'waves' is becomes urgent. In other words, the scientific realist seeks to provide an *interpretation* of quantum mechanics. Let me state the realist vantage point in Putnam's own words:

> According to [operationalism,] statements about [certain magnitudes, such as distance, charge, mass,] are mere shorthand for statements about the results of measuring operations... I shall here assume that this philosophy of physics is *false*... According to me, the correct view is that when the physicist talks about electrical charge, he is talking quite simply about a certain magnitude that we can distinguish from others partly by its 'formal' properties..., partly by the structure of the system of laws this magnitude obeys..., and partly by its *effects*. We know that [a charge meter] measures electrical charge, *not* because we have adopted a 'convention', or a 'definition of electrical charge in terms of meter readings', but because we have accepted a body of theory that includes a *description of the meter itself in the language of the scientific theory*. And *it follows from the theory* [...] that the meter measures electrical charge..." (1965, 130f; emphases in original)

If we accept this realist vantage point as the correct way of thinking about not just Newtonian physics, but also about quantum mechanics, then, Putnam infers, "the term 'measurement' plays *no fundamental role in physical theory as such*", i.e., " 'measurement' can never be an *undefined* term in a satisfactory physical theory…" (132; emphases in original) The implied rejection of the fundamentality of measurement, and hence of observers, plays a central role in Putnam's critical evaluation of the various proposed interpretations or proto-interpretations, as we will see shortly.

The question before us, then, is whether we can understand or interpret quantum mechanics in a way compatible with scientific realism, and if so, *how*? For the purposes of answering this question, Putnam analyzes the most visible interpretations at the time of writing in the two essays I shall be centrally concerned with here, viz. those two essays of forty years apart treating his eponymous looking at quantum mechanics. These interpretations include, for the 1965 essay, (a) de Broglie's pilot wave interpretation, (b) the "original" Born interpretation, (c) hidden-variables interpretations, and (d) the Copenhagen interpretation; for the 2005 sequel, Putnam considers (d), and *da capo*, (the von Neumann version of) the Copenhagen

---

starting points. The *Synthese* article of 1974 gives classical expression to Putnam's temporary advocacy of quantum logic, and two articles in the collection edited by De Caro and Macarthur (2012a, 2012b) serve well as recent points of contact.

interpretation, (e) GRW, (f) the many-worlds interpretation, and finally (g) Bohmian mechanics.

In the next section, Section 2, I shall consider a challenge to Putnam's claim that quantum mechanics stands in need of interpretation aired by Nancy Cartwright (2005) and argue that her objections do not succeed. Section 3 analyzes Putnam's own assessment of the various interpretative options that he discusses in 1965 and in 2005. Section 4 concludes with a brief summary of my main claims and emphasizes the historical importance of Putnam's work on the foundations of quantum mechanics.

## 2 Cartwright's challenge

Before we delve into the minutiae of Putnam's analysis of the various interpretations of quantum mechanics, we are well advised to pause over a dissent penned by Nancy Cartwright (2005). Her remonstrance is two-pronged. First, she argues that Putnam's claim of 1965 that quantum mechanics needs an interpretation is inconsistent, or at least stands in tension, with his 1962 work on the meaning of theoretical terms. Secondly, she insists, *pace* Putnam, that quantum mechanics does not need an interpretation. Let us consider these two objections in turn.

### 2.1 The tension with Putnam's 1962 work on theoretical terms

After reminding the reader of the seminal influence Putnam's 1965 piece exerted on the philosophy of quantum mechanics, Cartwright rushes to agree with its central conclusion, viz. that none of the extant interpretations of quantum mechanics can be deemed satisfactory. This situation, Cartwright proceeds, has not improved in the forty years hence, for contemporary attempts suffer from the very same kinds of difficulties that Putnam listed. Less than a page into the essay, however, the gloves come down. Despite the negative conclusion, Cartwright contends, we need not despair as the absence of an agreeable interpretation can easily be explained: there is no such interpretation because none is needed. In fact, she continues, this is the stance that Putnam would have been well advised to take in 1965, as it was he himself who argued—three years earlier and in a different context—that "*successful theories do not need interpretation*." (Cartwright 2005, 188; emphasis in original) More specifically, Putnam's claim that quantum mechanics needs an interpretation, and in particular that the quantum state, or the wave function, does, stands in conflict with what Cartwright takes to be the central conclusion of his famous paper 'What theories are not' of 1962, viz. "that theoretical terms do not need interpretation." (ibid., 192). This is because

> The theory itself, the entire theory with all its diverse uses and implications, gives meaning to the quantum state. A concept from a theory like this comes already interpreted. This doctrine about meaning begins in 'What Theories Are Not', where Putnam attacks the idea that scientific concepts should ever be in need of interpretation. (ibid.)

In 'What theories are not', Putnam assails what he takes to be the received view of theories in empirical science, viz. Rudolf Carnap's conception of a scientific theory as *partially interpreted calculus*, i.e.,

> as an axiomatic system which may be thought of as initially uninterpreted, and which gains 'empirical meaning' as a result of a specification of meaning *for the observation terms alone*. A kind of partial meaning is then thought of as drawn up to the theoretical terms, by osmosis, at it were. (Putnam 1962, 216; emphasis in original)

This conception of theories of course depends on an alleged dichotomy between 'observation' on the one hand, and 'theory' on the other. This dichotomy manifests itself in two forms. First, the non-logical vocabulary of the empirical science at stake is partitioned into 'observational' terms, such as 'green', 'hot', 'click', and 'theoretical' terms, such as 'boson', 'gene', or 'supply'. Secondly, the statements of an empirical science are then divided into 'observational' statements and 'theoretical' statements. The latter class comprises all and only statements which contain theoretical terms, while the former consists of statements which contain only observational terms (and logical vocabulary) and are thus free of theoretical terms.

Putnam then spends the remainder of the essay to argue that both forms of the dichotomy collapse under the weight of counterexamples and are thus not ultimately tenable. However, this seemingly catastrophic failure, he appeases the reader, engenders no alarming consequences for our philosophical project of understanding empirical science. Putnam contends that the main motivation for positing the dichotomy is flawed: the dichotomy was introduced to show how the justification of a theory results from the justification we may have for its 'observational basis', such that the theoretical terms acquire their meaning ultimately in purely observational terms. But the idea that justification necessarily originates in pure observation and then percolates from there to the (ever more) theoretical parts of science, Putnam expounds, cannot be maintained; rather, "justification in science proceeds in any direction that may be handy" (216).[2] Furthermore, both theory-observation distinctions introduced above are, Putnam complains, "completely broken-backed" (ibid.).

In order to substantiate this last claim, Putnam offers four sub-theses, each supported by examples. First, (excluding constructions such as conjoining every observational term with 'and is an observable thing' whose sole purpose is that of trivially applying only to observable things,) 'observation terms' are applicable to unobservable objects, a possibility at least neglected by Carnap. In fact, there is not a single term, it seems, that could not possibly be applied to unobservable entities. If 'observational term' was thus construed as to only apply to observables, then there are no 'observational' terms! Once it is granted, however, that 'observational' terms may apply to unobservables, Putnam contends, then the mystery of how terms referring to unobservables get introduced into a language evaporates. Furthermore, there is no longer any reason to assume that theories (and hence statements) about unobservables must contain 'theoretical' terms.

[2] In fact, Putnam also states the further contention that the alleged "*problem* for which this dichotomy was invented… does not exist." (ibid.) The problem, supposedly, was to account for the possibility to interpret theoretical terms at all. It is not clear to me, however, how there could be such a problem in the first place without there being some prior distinction between theoretical and non-theoretical terms. As suggested by the remarks following the cited passage, Putnam accepts that there is indeed such a pre-existing distinction, however imprecise, that the then sharpened dichotomy tries, but fails, to precisify.

The second sub-thesis asserts that "[m]any terms that refer primarily to what Carnap would class as 'unobservables' are not theoretical terms; and at least some theoretical terms refer primarily to observables." (217) Concerning the latter part, just consider how theory-laden terms like 'satellite' may be while still referring to observables; regarding the former, if we were to demand that all terms referring to unobservables are 'theoretical', then arguably that would include many terms of ordinary language ('angry' and 'loves' are Putnam's examples).

The demarcation between theoretical and observational *statements* is also problematic, as the third and fourth of the sub-theses illustrate. If we insist, with Carnap, that observational statements cannot contain theoretical terms, then many observational reports would be classified as 'theoretical' statements—just consider the report, reaching us from CERN, that 'We have observed the Higgs boson.' Fourth and last, there exist, Putnam claims, scientific theories which only refer to observables. Putnam's example here is Darwin's theory of evolution in its original form. This seems flatly false: terms like 'inheritance', 'species', 'selection', 'climate', and 'adaptive' seem hardly more observable than 'angry' and 'loves'. In fact, I cannot think of a scientific theory that refers only to observables. Be that as it may, Putnam takes both of these claims as evidence that no distinction between observational reports and theoretical statements can be based on the vocabulary of the relevant language alone. Neither should it, he adds.

Given that no principled distinction can be drawn between either theoretical and observational terms or between observational and theoretical statements, it is hardly surprising that Putnam denies that the theoretical terms (or the 'theoretical' ones), unlike observational terms, are only partially interpreted. He then considers, and rejects as either inadequate for Carnap's purposes or as incompatible with a weak form of scientific realism, three different ways of conceiving of 'partial interpretation' in the light of the failed precisified double dichotomy.

This leaves open the problem of how theoretical terms in science do, in fact, acquire meaning. But why, asks Putnam, think in the first place that an explication of the meaning of theoretical terms in purely observational ones is possible? Once we abandon Carnap's foundationalist empiricism for a minimally realist and more holist view of science, the problem loses much of its urgency. By thus exposing the foundationalist prejudices inherent in this formulation of the Carnapian problem, Putnam is then free to argue that theoretical terms are introduced and learned in roughly the same way as most other words.

Cartwright takes herself to be applying this very lesson to quantum mechanics when she states that its theoretical terms, such as 'wave function', do not

> need some special kind of interpretation… My claim here is that there is nothing peculiar about the quantum-state function. It is just like any other theoretical term, and Putnam's own conclusions apply to it. It does not need an interpretation, much less an interpretation in observational/classical terms; nor is such an interpretation likely to be possible. (2005, 197)

Instead, Cartwright continues, quantum mechanics receives all the 'interpretation' it needs through the myriad ways in which it is used and applied; it gets the "attachment [to the world] through experiment, technology, explanation and prediction." (ibid., 189) In other words, Cartwright seems to read Putnam to entail that if a 'partial interpretation' of the wave function in purely observational terms is not feasible (or even desirable) because, to put it positively, of the way in which theories get 'interpreted' through their use and application, then quantum mechanics does not stand in need of an interpretation, i.e., its formal apparatus needs no explicit tethering to the world.

The problem with this thesis—Cartwright's first—is that the measurement problem in quantum mechanics, to be introduced momentarily, does not at all rely on the empiricist foundationalism that Putnam attacks in 'What theories are not', and neither does it depend on the theory-observation dichotomy. But the measurement problem establishes, conclusively in my view, that quantum mechanics *does* need an interpretation. The measurement problem, and hence the need for interpretation, is not touched by Putnam's arguments of 1962. Consequently, there is no conflict between the 1962 claims and those of 1965—at least not if the measurement problem is understood as I will present it in §2.3 below. Arguably, this is not how Cartwright conceives of it; but if it captures (the correct precisification of) how Putnam conceives of it, then Cartwright's charge of inconsistency does not stick.

Not only are thus the concerns of 1962 and those of 1965 somewhat orthogonal to one another, but Putnam in fact explicitly leaves open the possibility that there may well be *some* theoretical terms in need of interpretation when he writes, admittedly parenthetically, but in the very same 1962 essay, that "giving the meaning of theoretical terms in science… may, of course, be a problem in specific cases." (225) I do not know whether Putnam wrote this remark with quantum mechanics in mind; but I wouldn't be surprised if he did. His earliest publication on quantum mechanics that I am aware of is a short commentary on the EPR paradox that was published in *Philosophy of Science* in 1961 (see bibliography). Furthermore, in 2005, he tells the reader that he has extensive conversations about the measurement problem with a world-famous physicist *in 1962*. Both of these facts bear witness that he must have been thinking about the foundations of quantum mechanics around the time he composed 'What theories are not'. If there were indeed an inconsistency, Putnam would have had to be a masterful compartmentalizer!

### 2.2 Cartwright on the measurement problem

As claimed above, Cartwright's contention is bipartite: not only does she think that Putnam's 1965 claims regarding quantum mechanics stand in awkward tension with his own 1962 rebuttal of Carnap's insistence on the 'partial interpretation' of theoretical terms, but she also maintains that he was mistaken to spurn, allegedly, his earlier work and avow that quantum mechanics has a serious interpretational issue. In other words, she maintains that no interpretation of quantum mechanics is needed, *for exactly the reasons given in Putnam 1962*. Central terms like 'wave function' acquire their meaning, and all there is to their meaning, by the way the quantum state behaves in all the many experimental settings in which quantum mechanics is tested, all the many technological applications in which it is used, and, of course, through the operative "constraints of theoretical practice"[3], i.e., through the full network of pertinent inferences—including theoretical ones—in which quantum mechanics is engaged. It is evident, she argues, that if we look at all these cases of experimentation and application, that "there is no single formula that covers all the various connections we find." (198) There is no "one magic formula" (ibid.) that would fit the bill, she continues, let alone one in purely observational terms.[4]

In §5, Cartwright discusses, and dismisses, the most famous of all arguments to the conclusion that there is a measurement problem in quantum mechanics: the thought experiment involving Schrödinger's cat. She discounts Schrödinger's cat as a "fantasy" (199) trading in unreal, and unrealistic, physical set-ups and processes. She complains that "of course there are no such quantum states as 'cat alive' and 'cat dead'", and that "[t]here is no such thing as the 'cat-alive/cat-dead' operator." (ibid.)

But this is altogether too quick: although it's true that Schrödinger's cat is a "story invented to match an abstract piece of formalism and an abstract concept of measurement" (ibid.), it points precisely to a seemingly insuperable obstacle in attaching quantum mechanics to the real world—regardless of how successfully the theory is used and applied! In other words, there really is a 'measurement problem', independently of how realistic Schrödinger's set-up with the cat and the various contraptions it involves is.

Before I move to restate the argument establishing that there is a measurement problem, it should be noted that Cartwright's own views on the measurement problem have evolved considerably since the beginning of her career. In her 1975, she was moved to accept that there is a measurement problem, and found realist interpretations denying the eigenstate-eigenvalue link such as that proposed by Danieri, Loinger, and Prosperi the most congenial and promising—against what she cites in 1983 as objections made in Putnam 1965![5] Putnam's objection, accepted by Cartwright in 1983 (cf. p. 170), was that the proposal fails to solve the measurement problem because the post-measurement state is still a superposition rather than an eigenstate. By 1983, thus, she rejects her earlier, overly realist, solution, in favour of a collapse interpretation asserting the 'reduction of the wave packet', trading on indeterministic "transition probabilities".[6] She denies, however, that there is a measurement problem; or, rather, she insists that there is no *measurement* problem, but instead the remaining issue of adjudicating when a system's dynamics is governed by the Schrödinger evolution and when a reduction of the wave packet occurs (see below). She insists, however, that his "characterization" problem is merely an artifact of the mathematical representation of the system's dynamics in terms of two mutually exclusive categories of evolutions when in fact the real physical system just evolves according to the more general 'quantum statistical law' encompassing both. As this more general dynamical law encompasses both unitary Schrödinger evolution as well as non-unitary dynamics, if one cuts through her rhetoric, then rather than denying the measurement problem, she can be read as advocating a solution to it based on a rejection of what will be claim (B) in the next section (§2.3). But I shall desist from any sustained assessment of her 1983 proposal and only add that I concur with Butterfield's (1983) conclusion that Cartwright offers more of a research agenda than a full-fledged theory.

[3] Cartwright, private communication on 11 June 2014.

[4] One might worry that this mischaracterizes the problem, which is not to give the meaning of the term 'wavefunction', and perhaps of other theoretical terms, in classical or purely observational terms, but instead to parse out what the theory asserts regarding the aspects of the world it seeks to describe. But what Cartwright perhaps has in mind here is the distinction between the microscopic and the macroscopic world on which traditional attempts to solve the measurement problem rely. Most contemporary approaches to solve the measurement problem, however, are 'observer-independent', i.e., they assume no such fundamental dichotomy between quantum systems and classical observers.

By 1999, Cartwright moved away from her collapse approach awarding centre stage to transition probabilities in order to endorse what she dubs a "have-your-cake-and-eat-it-too" strategy to address the "problem of superposition" (1999, 212). Even though there were inklings of an anti-fundamentalist stance earlier, she now (in 1999) identifies as the enemy not the realist, but instead the fundamentalist, who takes quantum mechanics (or anyway some quantum theory or other) to be a fundamental und universally applicable theory. Against quantum fundamentalism, Cartwright contends that "quantum theory is severely limited in its scope of application." (ibid., 214) In a dappled world, macroscopic measuring devices must be described by a different theory from the electrons with which they interact. In the classical theory of the macroscopic objects, no superposition states are allowed. Classical and quantum mechanics are both necessary, and neither is sufficient by itself. Consequently, there simply is no mystery as to why measurement outcomes are determinate. This raises the obvious question of what governs interactions between the two realms. For Cartwright, the answer must be found in the many connections forged in actual scientific, and particularly experimental, practice, and not in some unique magic formula relating the two. Cartwright acknowledges that there is no guarantee that this 'retail' solution will never run into an inconsistency between the two descriptions; this possibility, however, is a far cry from the universal angst spread by the scaremongers of 'wholesale' quantum imperialism. She insists that "[s]o far we have failed to find a single contradiction" (233) between classical and quantum physics, and challenges the reader to identify an inconsistency. Unfortunately, Cartwright's confidence is misplaced, as such an inconsistency can not only be identified, but is found to be both rampant and pervasive. This is the topic of the next

---

[5] Putnam (1965) does not explicitly mention the proposal due to Danieri et al.

[6] Cf. 1983, Ch. 9.

subsection.

### 2.3 Does quantum mechanics need an interpretation?

Let me briefly summarize the main tenets of the theory to see how the measurement problem arises.[7] In fact, we only need three basic principles of quantum mechanics to derive the problem. The first asserts the completeness of the quantum-mechanical description of the state of a system:

> **Principle 1 [Completeness]**
> The wave function contains the complete information about the physical state of the system at stake.

In fact, the state of a quantum-mechanical system is more generally described by a unit vector in a so-called *Hilbert space*, i.e., a vector space with an inner product.[8] The reason why vector spaces turn out to be so perfectly adequate for the task of representing the state spaces of quantum-mechanical systems is because they naturally encode another principle of quantum mechanics, the infamous superposition principle:

> **Principle 2 [Superposition]**
> If $|A\rangle$ is a possible physical state (or the wave function describing it) of some system, and if $|B\rangle$ is a (distinct) possible physical state of the same system, then $\alpha|A\rangle + \beta|B\rangle$ is also a possible state, where $\alpha, \beta \in \mathbb{C}$.

Here, as below, I use Dirac's 'bra-ket notation' of denoting vectors in a Hilbert space by '$|A\rangle$', '$|B\rangle$', etc. Principle 2 is notable because it is clearly not valid in classical physics: if a system can be located *here*, and if it can also be located *there*, it does not follow that it can simultaneously be located *'here and there'*, whatever exactly that may mean. Linear combinations of admissible states with classically admissible properties will not in general also be admissible states with classically admissible properties. Principle 2 asserts that this is *always* possible in quantum mechanics, for any pair of admissible states.

Since the measurement problem arises dynamically, we also need a principle concerning the dynamical evolution of quantum-mechanical states. Here it is:

> **Principle 3 [Schrödinger evolution]**

---

[7] For a very accessible introduction to quantum mechanics, its measurement problem, and many of the various proposed solutions, cf. Albert 1992. A particularly sharp formulation of the measurement problem has been given in Maudlin 1995. The presentation of the problem here owes much to both of these sources.

[8] More precisely, unit vectors (or rays) in a Hilbert space represent *pure states*. Thus, a further generalization would include mixed states as well. However, this generalization is unnecessary for the point to be made here.

> Given the state of any physical system at any 'initial' time, and given the forces and constraints to which the system is subject, the *Schrödinger equation* gives a prescription whereby the state of that system at any other time is uniquely determined.

This dynamics of the state vector is thus *deterministic*. You may be aware of the standard lore according to which quantum mechanics is indeterministic. Whether this is indeed so is actually a contentious matter, and depends on one's solution of the measurement problem, as we will see below, and consequently, on the interpretation of quantum mechanics one endorses (Wüthrich 2011). What is uncontroversial however, is that *if* Principle 3 states everything there is to be stated about the dynamical evolution of a quantum-mechanical state, as we assume for the moment, then that evolution is deterministic.

There is no need for us to consider the full mathematical glory of the Schrödinger equation; in fact, the only property of the equation relevant for the purposes of incurring a 'measurement' problem is that it is linear. A dynamical law is *linear* just in case if any state $|A\rangle$ at time $t_1$ is evolved into another state $|A'\rangle$ at a later time $t_2$ and any other state $|B\rangle$ at $t_1$ is evolved into $|B'\rangle$ at $t_2$, then $\alpha|A\rangle + \beta|B\rangle$ at $t_1$ is evolved into $\alpha|A'\rangle + \beta|B'\rangle$ at $t_2$. In other words, a superposition evolves just as a superposition of the evolutes of its component states. What this means, of course, is that whenever a system starts out in a superposition state, it will always remain in a superposition state for as long as Principle 3 alone governs its dynamical evolution.

Suppose now that we have a quantum system with a spin degree of freedom, such as an electron, and a measuring device equipped to measure the spin in a particular direction orthogonal to the direction of flight of incoming particles. In order to keep things simple, let us just assume that it has a dial with three settings, 'ready', '$z$-up', and '$z$-down', indicating that the device is ready to make a measurement, has found the particle to have up spin in the $z$-direction, and down spin in the same direction, respectively. We start the experiment by preparing the device to be in the ready state (such that it reads 'ready'), and then feed electrons into it in order to get their $z$-spin measured. The measurements are then recorded by the final position of the dial ('$z$-up' or '$z$-down', as the case may be).

It is important to assume that the measurement apparatus measures 'faithfully', i.e., that it 'correctly' records the spin of the measured electrons. More precisely, we demand that

$$|\text{ready}\rangle_m |z\text{-up}\rangle_e \rightarrow |\text{'}z\text{-up'}\rangle_m |z\text{-up}\rangle_e \qquad (1)$$

$$|\text{ready}\rangle_m |z\text{-down}\rangle_e \rightarrow |\text{'}z\text{-down'}\rangle_m |z\text{-down}\rangle_e \qquad (2)$$

where the subscripts $m$ and $e$ designate the states of the measuring device and the electron, respectively. Note the quotes in the post-measurement state of the device; they should indicate that the device is not itself in a $z$-up or $z$-down state, but rather in the state of having recorded a $z$-up or $z$-down state.

From Principle 2, (1) and (2), and from the linearity of the dynamical law of Principle 3, it follows that there exists an '$x$-up' state for the electron,

$$|x\text{-up}\rangle = \frac{1}{\sqrt{2}}|z\text{-up}\rangle + \frac{1}{\sqrt{2}}|z\text{-down}\rangle,$$

which, if fed into a ready measuring device, evolves, with certainty, into the total state

$$\frac{1}{\sqrt{2}}|\text{'}z\text{-up'}\rangle_m |z\text{-up}\rangle_e + \frac{1}{\sqrt{2}}|\text{'}z\text{-down'}\rangle_m |z\text{-down}\rangle_e. \qquad (3)$$

The problem with (3) is that the dial of the device is in a superposition of showing that the electron had spin up and that it had spin down. But however probabilistic the dynamics may be, measurements seem to have determinate outcomes, so that in the present case we ought to get for the total state of the device-cum-electron *either* $|\text{'}z\text{-up'}\rangle_m |z\text{-up}\rangle_e$, with 50% probability, *or* $|\text{'}z\text{-down'}\rangle_m |z\text{-down}\rangle_e$, with 50% probability, rather than some superposition state. Although it violates Principle (3), this leads to states that permit definite measurement outcomes. But it is measurably[9] different from (3)! On the other hand, (3) is a state in which there is simply no fact of the matter about where the pointer is pointing.

This leads to the following sharp formulation of the measurement problem, which I borrow from Maudlin 1995, and which has now become standard:

> The following three claims are mutually inconsistent.
>
> [(A)] The wave-function of a system is *complete*, i.e. the wave-function specifies (directly or indirectly) all of the physical properties of a system.
>
> [(B)] The wave-function always evolves in accord with a linear dynamical equation (e.g. the Schrödinger equation).
>
> [(C)] Measurements of, e.g., the spin of an electron always (or at least usually) have determinate outcomes, i.e., at the end of the measurement the measuring device is either in a state which indicates spin up (and not down) or spin down (and not up). (1995, 7)

Since the three propositions (A), (B), and (C) are mutually inconsistent, at least one of them has to be given up. But simply pointing out one of them as to be denied does not by itself constitute a *solution* to the measurement problem: since denying any of them either strikes down, or at least seriously curtails, a major principle of quantum mechanics (in the case of (A) and (B)) or else invalidates what appears to be a perfectly uncontroversial assertion concerning our experimental practice (in the case of (C)), one must complement the rejection of a chosen proposition with a *theory* which delivers a complete way of thinking about the quantum-mechanical system and how it relates to our observable world.

Solutions to the measurement problem can naturally be classified, then, in accordance to which of the propositions above they deny. Hidden-variables theories reject that the wavefunction captures all the physical degrees of freedom of a quantum system and thus deny (A). Theories in this camp have to list what these additional degrees of freedom are, and specify the dynamical equations which govern their evolution, including how they couple to the wavefunction. The most prominent theory denying (A) is, of course, Bohmian mechanics.

---

[9] Albert 1992, p. 76n. Cf. also ibid., Ch. 5 (particularly the digression).

So-called 'collapse theories', such as GRW, amend the linear evolution asserted in (B) with episodes of non-linear evolution that explicitly violate the Schrödinger equation, or replace the latter altogether with a more general—and non-linear—dynamical law. The central task of approaches in this family is to give the full non-linear dynamics, and to lay down exactly under which conditions it applies.

The final group of solutions repudiates (C); these so-called 'many-worlds' theories aver that we only seemingly obtain determinate measurement outcomes when in reality, the state of the total system of both electron and measuring device is still as in (3). Approaches disavowing (C) must then give an account consistent with the quantum-mechanical statistics we experimentally find of why measurement outcomes appear determinate when in fact they are not.

Whatever the merits and demerits of the different approaches to solving the measurement problem may be (we will get to that shortly), let me close the section by emphasizing the two conclusions we have reached so far. First, demanding an interpretation of quantum mechanics is perfectly consistent with rejecting empiricist foundationalism and its attendant theory-observation dichotomy. Second, the measurement problem offers an inconsistency that must be addressed; any solution to it will, ipso facto, offer an interpretation of quantum mechanics.

## 3 Putnam's assessment of the interpretative options

The central purpose of Putnam's two articles of 1965 and 2005 is the same: to evaluate some of the promising, or anyway most famous, interpretations of quantum mechanics that have been suggested. In 1965, these include most prominently de Broglie's pilot wave interpretation, Born's "original" ignorance interpretation, hidden-variables theories, and the Copenhagen interpretation. Forty years later, the focus is on collapse theories (some of which did not yet exist in 1965), many-worlds interpretations, and hidden-variables theories again. The goal of the present section is to go through at least some of Putnam's evaluations of the various approaches, so that the reader gets a sense of how incisive and (mostly) apt Putnam's criticisms of the discussed programs are. Given the scope of this article, I cannot possibly do full justice to either these programs or to Putnam's criticisms of them. The necessary brevity of my remarks may thus occasionally puzzle those with no prior familiarity with interpretations of quantum mechanics. Where appropriate, I will try to make amends by giving references to work that can fill the gaps I leave open.

### 3.1 Taking a first look: 1965

Putnam's 1965 essay "A philosopher looks at quantum mechanics" starts out from an assertion of scientific realism, as explicated in Section 1. Putnam takes the rejection of the fundamentality of measurement to be a straight consequence of his realism. This means that 'measurement' cannot function as a primitive term in any theory claiming to solve the measurement problem.

Starting out from the fact that the quantum-mechanical state can be represented by a wavefunction, Putnam then addresses the question of the significance of these 'waves'—a central issue for any solution of the measurement problem.[10] The first answer Putnam considers was also the first historically: de Broglie's 'pilot wave interpretation'. De Broglie formulated the pilot wave approach in 1927, one year after Schrödinger published his wave approach to quantum mechanics. The approach was quickly abandoned, but is historically significant as a stepping stone toward Bohmian mechanics.[11]

De Broglie's main idea was simply to reify the 'waves' used to represent quantum-mechanical states: rather than mere representations, the waves are physical entities, so that the state of the system simply *is* the system of waves. Putnam unceremoniously—and rightly—discards this approach as unsuccessful. The first problem to note is that quantum 'waves' are rather unlike the waves we know and love. For starters, unlike real waves, they have complex amplitudes. Moreover, they live in Hilbert space, which is *not* physical space, but the mathematically abstract 'state space' of quantum mechanics. Finally, the 'reduction of the wave packet' or collapse seemingly necessary to reproduce the experimental outcomes actually found subjects the physical wave to a nonlinear (Putnam says "discontinuous", but this need not be so) change upon measurement from being very spread out to a highly localized wave packet. At the very least, such momentary contractions seem incompatible with the usual behaviour of physical waves, thus considerably lessening the appeal of an interpretation of the wave function in terms of physical waves.

The second interpretation analyzed by Putnam is Born's "original" interpretation, according to which the elementary particles described by quantum mechanics are *classical*, insofar as they are pointlike masses with determinate position and velocity. However, these particles don't obey the classical laws of Newtonian mechanics. Instead, they are ultimately governed by quantum-mechanical laws. On Born's interpretation, the wave function does not represent the state of the system, but rather of our incomplete knowledge of it, such that the square of the intensity of the wave inside a given region gives the probability that the particle is actually there.[12] This way of conceiving of the wave function's meaning avoids the difficulties that de Broglie's pilot wave interpretation faced, in particular vis-à-vis collapse, when it insisted on the physicality of the 'wave': as it merely represents our knowledge, we seem free to interpret it as a purely mathematical object tracking what we know about the state of the system at stake and the sudden jolts that occur the instant we learn of a measurement outcome. However, just as we evade the troubles that collapse foisted on any attempt to take the wave as a physical object, we lose its ability to offer a physical mechanism that accounts for the physical interference patterns that we observe on the detection screen in two-slit experiments. How can such interference occur if there are no waves that produce it?

[10] And the topic of a recent collection of essays, cf. Ney and Albert 2013.
[11] For the connection to Bohmian mechanics, see Goldstein 2013.
[12] This is of course 'Born's Rule', which is taken over e.g. by the later 'Copenhagen interpretation', which is indebted to Born's "original" interpretation.

More decisively, Putnam shows how Born's interpretation cannot adequately deal with superpositions. Suppose we have a large ensemble of quantum systems such that each of them can have any of two incompatible properties $\mathcal{A}$ and $\mathcal{B}$. These properties are associated with operators $\hat{A}$ and $\hat{B}$ on the single-system Hilbert space, and these operators have eigenvectors in that Hilbert space such that $\hat{A}|a\rangle = \alpha|a\rangle$ and $\hat{B}|b\rangle = \beta|b\rangle$. Assuming that we have available a perfectly faithful measurement apparatus, we will thus obtain the following measurement outcomes: if all the systems in the ensemble are in state $|a\rangle$ (or $|b\rangle$, respectively), then 100% of the measurements on single quantum systems will yield outcome $\alpha$ (or $\beta$, respectively). According to Principle 2 above, the state $|c\rangle = \lambda_1|a\rangle + \lambda_2|b\rangle$ is also a possible state of an individual quantum system in the ensemble. Measuring a number of systems prepared to be in the identical state $|c\rangle$ we might obtain that

(a) measurements of observable $\mathcal{A}$ yield outcome $\alpha$ in 60% of the cases,

but we might also get that

(b) measurements of observable $\mathcal{B}$ yield outcome $\beta$ in 60% of the cases.

Since the properties $\mathcal{A}$ and $\mathcal{B}$ are incompatible, i.e. $[\hat{A}, \hat{B}] \neq 0$, there are no simultaneous eigenstates for $\hat{A}$ and $\hat{B}$, and we cannot perform simultaneous measurements of both properties. In other words, if we first measure one of them on some systems, and then measure the other *on the same systems*, then we will obtain different statistics than if we performed only the second type of measurements. Thus, we cannot make measurements to check (a) and (b) on the same systems. However, if the wave function or the quantum state merely represented our incomplete knowledge of the systems' states—rather than some physical feature of these systems—, it would be utterly mysterious how our knowledge could be affected by the order of measurements on otherwise identical systems. The only explanation seems to be that the measurements we perform make a physical difference; they change the systems' states rather than just our knowledge of them.

Putnam identifies the following principle to which Born's interpretation is committed as the culprit:

> **Principle [No Disturbance (ND)]**
> "The measurement does not disturb the observable measured—i.e. the observable has almost the same value an instant before the measurement as it does at the moment the measurement is taken." (138)

*Before* any measurement is made on the systems prepared to be in state $|c\rangle$, (a) and (b) cannot *both* be true—but they would have to be if ND were true. Thus, experimental facts concerning quantum systems such as those described above seem plainly incompatible with ND.[13] Given the overwhelming evidence conflicting with it, ND ought to be rejected, and with it Born's original interpretation.

Next, Putnam turns to so-called 'hidden-variables interpretations', i.e. interpretations that deny proposition (A) in Maudlin's articulation of the measurement problem. The front runner in this camp is (and was in 1965) Bohm's mechanics, originally proposed in Bohm 1952.[14] Bohmian mechanics takes the fundamental ontology to consist of particles, which all have determinate positions. ND is valid only for position measurements. Other properties these particles may have, such as spin, are only contextually exemplified and thus not fundamental, and hence ND is false of them. Putnam's main criticism of hidden-variables theories is that the (general) falsity of ND forces them to adopt very strange laws. For instance, Putnam complains, Bohm's theory requires a strangely behaving 'quantum potential', i.e. an unknown force for which there is no (direct) evidence, in order to account for the violation of ND brought about by the disturbance of the state by the measurement. This criticism is somewhat misguided—at least by present lights, as most Bohmians no longer defend the original 'quantum potential' version of their theory and instead regard it as a "first-order theory, in which it is the velocity, the rate of change of position, that is fundamental." (Goldstein 2013, Sec. 5)

Putnam proposes that we accept the general invalidity of the principle ND, including for position measurements, as a "condition of adequacy" (145) any candidate interpretation of quantum mechanics must satisfy.[15] This would of course rule out Bohmian mechanics. But

---

[13] See Albert 1992, Chapter 1, for a very accessible account of why measurements on quantum systems which are not in an eigenstate of the measured observable necessarily amount to a 'disturbance', at least if the quantum state is taken to be complete and measurement outcomes determinate.

[14] For the most up-to-date and authoritative account of Bohmian mechanics, see Goldstein 2013. Cf. also Albert 1992, Chapter 7.

[15] Putnam's three conditions of adequacy are as follows:
"A. The principle ND should not be assumed even for position measurement.
"B. The symmetry of quantum mechanics, represented by the fact that one 'representation' has no more and no less physical significance than any other, should not be broken. In particular, we should not treat the waves employed in one representation (position representation in the case of the hidden variable theorists) as descriptions of physically real waves in ordinary space.
"C. The phenomena of superposition of states... must be explained in a unitary way." (145f)
Putnam claims that hidden-variables interpretations stand in violation of all these conditions. Bohmians would naturally repudiate conditions A and B—their violation, they will rightly insist, is a *feature*, not a bug of their interpretation. Bohmian mechanics does have its issues, but they lie elsewhere. I am not sure about the precise meaning and status of condition C. Putnam charges Bohmian mechanics with a violation of C based on the now obsolete there is really no reason why we should accept such a sweeping a priori condition of adequacy. Thus, his argument against Bohmian mechanics does not ultimately succeed, as he acknowledges in 2005. But in 1965, he rejects hidden-variables interpretations as solutions to the measurement problem.

The fourth interpretation analyzed by Putnam is the Copenhagen interpretation, formerly also the *standard* interpretation of non-relativistic quantum mechanics. At core, the Copenhagen interpretation assumes the completeness of the quantum state, takes measurement outcomes to be determinate, and fundamentally distinguishes between physical systems that are 'measuring devices' and those that are not. It will thus have to deny (B) above. Roughly, then, interactions between a measuring device and a non-measuring system will lead to the 'collapse' of the quantum state of the latter, while interactions among non-measuring systems have no such effect.[16] This would solve the measurement problem if we are given a principled, and fundamental, distinction between the two kinds of physical systems, as well as a precise measurement dynamics, i.e. a dynamical law for those occasions, identified in a principled manner by an answer to the first request, when the usual Schrödinger evolution is upended. Unfortunately, the Copenhagen interpretation furnishes neither.

On the Copenhagen interpretation, particles do not have determinate properties, but only *propensities* to be unveiled by suitable measurements. This leads to a slight modification of Born's rule: the squared amplitude of the wave does not measure the probability that the particle *is* in a certain place, but instead the probability that it would be *found* in that place if a position measurement were made. This interpretation rejects ND by assuming 'complementarity', i.e. the idea that $\mathcal{A}$-measurements and $\mathcal{B}$-measurements cannot be made simultaneously.

Putnam judges that the Copenhagen interpretation straightforwardly satisfies his first two conditions of adequacy, but that the third is doubtful, since superpositions are not *explained*, but instead stipulated as primitives of the theory. More importantly, Putnam identifies the interpretation's reliance upon 'measurement' as a primitive term as its main deficiency because it violates one of the basic assumptions of scientific realism. The problem remains standing even 40 years after the inception of the theory, Putnam argues, because various attempts to spell out what 'measurements' amount to are unsuccessful due to the fact that they leave *other, related* terms unacceptably primitive, such as 'macro-observable'.

In effect, Putnam thus articulates what is now the standard complaint against the Copenhagen interpretation as I lodged it above: this leads directly to the measurement problem in the form of Schrödinger's cat and to the attendant difficulty of making sense of the dual requirements of separating the world into the measured and the measuring and of the demand of the universal applicability of quantum mechanics. According to the Copenhagen interpretation, micro-observables, i.e., those associated with the measured

---

reliance on a 'quantum potential'. Be that as it may, I find nothing intrinsically objectionable in the way modern Bohmians explain superpositions.

[16] For details of this view—skipped here—cf. Fay 2008.

system, don't exist unless they are in fact measured, but the macro-observables, i.e., those pertaining to the measuring system, take sharp values at all times. This renders measurements as an interaction between micro- and macro-observables, and it awards a special status to macro-observables insofar as they retain sharp values. This, Putnam criticizes, amounts to an underived assumption of the theory standing in tension with quantum mechanics' claim to universality. Putnam summarizes:

> The question we face is whether from such a quantum-mechanical characterization of a macro-observable together with the laws of quantum mechanics it is possible to deduce that macro-observables retain sharp values whether a measurement interaction involving them is going on or not. If we can do this, then the appearance of paradox and the *ad hoc* character of the [Copenhagen interpretation] will disappear. In spite of a number of very ingenious attempts, it does not appear that this can be done. (150f)

Putnam's claim in the last sentence of the quote is based on his ensuing discussion of how these attempts fail to circumvent the problem with Schrödinger's cat. The Copenhagen interpretation, Putnam infers, is thus at best an incomplete interpretation of quantum mechanics. Given his negative 1965 analysis of the four then main competitors to furnish such an interpretation, Putnam feels compelled to end on a rather bleak note: "In conclusion, then, *no* satisfactory interpretation of quantum mechanics exists today." (157; emphasis in original)

### 3.2 Taking another look: 2005

The general conclusion four decades later is only slightly less dreary. But before we get there, let me note some salient differences (and similarities) to the analysis Putnam offered in 1965. First, John von Neumann's version of the Copenhagen interpretation is rejected, and for similar reasons as was the version due to Werner Heisenberg and Niels Bohr forty years earlier. Second, the main argument of 1965 against hidden-variables theories is retracted. Third, new interpretations, largely unknown or altogether unarticulated in 1965, are being considered. Fourth, the focus shifts a bit from analyzing various interpretations—and rejecting them as inadequate—to classifying them. Fifth, a new condition of adequacy is introduced: 'Einstein's bed'. 'Einstein's bed' plays a central role in Putnam's rejection of von Neumann's collapse postulate, so let's turn to that first.

When Putnam joined the Princeton faculty in 1953, his PhD advisor[17] Hans Reichenbach organized for him to meet with Einstein. So when Einstein and Putnam met and when they talked about quantum mechanics, Einstein said something along the following lines:

> 'Look, I don't believe that when I am not in my bedroom my bed spreads out all over the room, and whenever I open the door and come in it jumps into the corner.' (624)

[17] And my own academic great-grandfather. I am honored to note, incidentally, that this makes Putnam my academic granduncle!

If we assume some mild form of realism and the universal validity of quantum mechanics and hence reject the possibility that our bed is in no determinate position but smeared across our bedroom when we are out looking at the moon, then we will share Einstein's disbelief. If we reject unobserved deliquescent beds and far-flung moons for these reasons, then we cannot grant measurements the exclusive power to 'collapse' quantum states and should join Einstein in repudiating von Neumann's collapse postulate which asserts just that. As Putnam puts it, the determinateness of measurement outcomes should naturally fall out of the theory—not by stipulating some ad hoc principles such as the collapse postulate.

Turning to the classificatory task then, Putnam distinguishes between two principal families of interpretations: those that invoke some sort of collapse and those that don't.[18] Both families of interpretations fall into two classes of their own, depending on their answer to the central question each family faces. For collapse theories, this means that they need to take a stance as to whether the collapse is induced by a measurement, i.e., by some physical system external to the measured system and not subject to the superposition principle, as von Neumann envisages, or whether collapse occurs in some other way. If we accept Einstein's bed as a condition of adequacy for any interpretation, then proposals in the quadrant populated by theories which take the collapse they accept to be prompted by measurement are out. The collapse theories that remain, then, are those which do not rely on measurements and are thus not 'observer-dependent'. The most important among them are 'spontaneous' collapse theories such as GRW, to which we will turn shortly.

For those interpretations which deprived themselves of the services of a collapse of the quantum state, however occasioned, to obtain determinate measurement outcomes, the central question before them is whether they permit additional degrees of freedom—hidden variables—that would determine the determinate measurement outcomes. Among those interpretations which accept hidden variables we find of course Bohmian mechanics. Here, the important thing for our present purpose is to note that Putnam (2005, 622f) retracts his earlier criticisms of Bohm's theory and insists that it cannot, at this point, be dismissed from consideration. Finally, any approach forfeiting the services of *both* collapse and hidden variables, as should be clear from the characterization of the measurement problem as given in Section 2.3 above, is forced to give up the seemingly natural tenet—(C) above—that measurement outcomes are in fact determinate. Everettian many-worlds interpretations fall into this camp.

Let us turn, with Putnam, to an analysis of GRW.[19] GRW and related collapse theories deny proposition (B) in §2.3 and solve the measurement problem by offering a new dynamical law. Since these theories are observer-independent, they do not dissect physical process into two fundamentally distinct kinds (ordinary processes and measurements), thus circumventing one difficulty of the Copenhagen interpretation. In GRW (and related theories), the collapse is not induced by a measurement but occurs spontaneously according to a modified dynamical law which includes stochastic and non-linear terms. According to

[18] See Table 2 on page 626 of Putnam 2005.
[19] GRW was proposed in Ghirardi, Rimini, and Weber 1986. For a systematic introduction with an eye to the philosophical foundations, cf. Ghirardi 2011.

GRW, each fundamental quantum system in the universe has a tiny, but non-zero probability that the system spontaneously localizes in some small region. Einstein's bed—let alone the moon—thus always has a determinate position, even if no one is near the bedroom. The reason for this is that even though each of the bed's particles only has a minuscule probability to be spontaneously localized in a location near a corner of Einstein's bedroom over the next few milliseconds, because the bed is made up of a very large number of particles, however, the probability that some particle is spontaneously localized during that interval is essentially one. And if one of the bed's particles undergoes a spontaneous localization, all the others get whacked into their place, courtesy of the bonds that obtain among the particles. So while it is *possible* that Einstein's bed smears across the bedroom, this is fantastically unlikely; so unlikely in fact that Putnam is safe to say that "it will never happen." (624)

Even though it keeps Einstein's bed in place, GRW is still faced with significant challenges, as recognized by Putnam. One of the major deficiencies of collapse theories, including GRW, is that they are not Lorentz invariant and thus seem not to afford a relativistic generalization that most believe is necessary. The tension between collapse theories and special relativity arises because the state that collapses is a global, i.e., spatially extended, state of a system at a given time and that appears to stand in stark contradiction to the relativity of simultaneity at the heart of relativity. Just as Bohmian mechanics, GRW is a non-relativistic theory. However, as John Bell recognized in 1987, GRW comes equipped with a property, 'multi-time translation invariance', that renders it as relativistic as a non-relativistic theory can get. Still, it is not relativistic, and this undercuts GRW's status as a fundamental theory. First steps to remedy that situation are undertaken in Tumulka 2006.

Putting concerns regarding Lorentz invariance to the side, it should be noted that GRW navigates a narrow passage between violating energy conservation and (allegedly) failing to solve the measurement problem. If the spontaneous localization of elementary particles were arbitrarily sharp, then, as Putnam remarks, its complementary magnitude, the particle's momentum, would be arbitrarily indeterminate courtesy to uncertainty, leading to possibly arbitrarily high energies, in violation of energy conversation. Of course, this threat is minimized in GRW in that the spontaneous localization is not fully sharp. This limits possible violations of the conservation of energy to unobservably small scales. However, even though the wave function is strongly peaked around one location, the wave function in general does not vanish elsewhere and in fact may be non-zero everywhere in space—even after a spontaneous localization. On the standard interpretation of the wave function, there is then a non-vanishing probability that, to put it coarsely, we find Einstein's bed in the patent office, not in his bedroom! This has become known as the 'problem of the tails' of the wave function. If this were the case, then even though the probability of Einstein's bed ending up anywhere but his bedroom is vanishingly small, this would hardly constitute a satisfactory solution to the measurement problem.

As Giancarlo Ghirardi (2011, §12) notes—quite correctly in my view—the problem of the tails only arises if one accepts the standard 'probabilistic' interpretation of the wave function, i.e. the one effectively captured by Born's rule. Arguably, it can be circumvented if one adopts a 'primitive ontology' according to which the wave function describes the mass density or the 'flash' density. On such an interpretation, the fact that the wave function does not vanish elsewhere does not mean that there is a small chance that Einstein's bed appears in the patent office, but rather that a tiny part of it in fact appears there.[20] This solution to the tails problem is not uncontroversial, but I refer the reader to Ghirardi 2011 (§12) and references therein for details of the controversy.

Despite these troubles, GRW remains on the table for Putnam. This is not the case for Everettian many-worlds interpretations.[21] Solutions to the measurement problem in this family deny (C), i.e., the idea that there are determinate measurement outcomes. Determinate outcomes of measurements, on this approach, are illusory; all possible measurement outcomes in fact obtain. Maximal sets of consistent measurement outcomes constitute the many 'worlds' there are according to these interpretations. The specific many-worlds interpretation that Putnam considers is the somewhat outdated version essentially due to Bryce DeWitt (1970) according to which the world physically splits into two each time a measurement with two possible outcomes is performed.[22] Let us see what Putnam thinks the problem with many-worlds interpretations is.

Suppose you are about to make a *z*-spin measurement on a particle in an *x*-spin eigenstate. As you perform the measurement, the world splits into two: in one world, the *z*-spin will be up, it will be measured to be up, and you will become aware that it is up; in the other, it will be down, it will be measured to be down, and you will become aware that it is down. The problem is that such an account, it seems, cannot, in general, get the quantum mechanical probabilities right. First, there is a sense in which it is simply false to attribute any probability to an outcome since all outcomes do in fact transpire, with probability 1. This *prima facie* problem can be evaded if we do not conceive of the quantum mechanical probabilities as pertaining to the sum total of existence, but rather as an experimentalist faced with the above set-up asking herself 'With what probability will I observe 'spin up'?'. Given that there are two possible outcomes and hence two post-measurement worlds, and given that the particle will be in a spin-up state in one of these worlds, following the principle of indifference it seems reasonable to say that the relevant probability is 0.5. It's an easy matter to extrapolate the probability of any of the possible combinations of outcomes for the case of many repeated measurements on identically prepared systems. As Putnam notes, it is just "simple combinatorics" (630). And it turns out that quantum mechanics is in perfect agreement with this way of calculating the probabilities.

So far, so good. But there is a second, much deeper problem. Just how serendipitous the previous case was can be seen if we now suppose that the outcomes of some measurements are not equiprobable, as they were above, but instead heavily biased toward one outcome. Say 'spin up' is much more likely than 'spin down'. The combinatorial probabilities will be the same as above, with a run of 'spin up' just as probable as an equally long run of 'spin down'. But this time, quantum mechanics will give very different probabilities, with a run of 'spin up' coming out as much more probable than an equally long run of 'spin down'. The experimenters in any world in which mostly 'spin up' was measured will correctly conclude that 'spin up' was much more likely and will see, suppose, quantum mechanics corroborated. However, the equally many experimenters in the equally many worlds of mostly 'spin down' will find the quantum-mechanical probabilities disconfirmed. Putnam then rightly asks how come we have been so incredibly lucky as to always have found the quantum-mechanical probabilities borne out in our world? In fact, on this interpretation, Putnam quips, quantum mechanics would be "the first physical theory to predict that *the observations of most observers will disconfirm the theory*." (630; emphasis in original) Thus, many-worlds interpretations are *empirically incoherent* in that their truth would undermine any (empirical) reason we could have to believe the theory to be true. On these grounds, Putnam rejects as incoherent solutions to the measurement problem in this family.

---

[20] 'Tiny' doesn't quite capture just how minuscule that part will be: Ghirardi (2011, §12) estimates that for a regular spherical ball of a mass of 1kg and a radius *r*, the mass that will be found outside a spherical region around the center of the ball with radius $r + 10^{-5}$ cm after the localization is of the order of 1 over 10 to the power of $10^{15}$ times the mass of a proton.

[21] For an introduction to many-worlds interpretations, cf. Vaidman 2014.

[22] Or into *n* worlds if there are *n* possible measurement outcomes. For a more contemporary take on many-worlds interpretations, cf. Wallace 2012.

This is not the place to defend many-worlds theories against this charge. It should be noted, however, that whether they are empirically incoherent in the sense just stated and—if so—they can recover Born's rule is a highly controversial matter and remains, as far as I am concerned, very much an open question.[23]

There is another—minor, but routinely misstated—point over which I would like my dissent recorded. Putnam writes that the many-worlds interpretations "give up the distinction between actuality and possibility" (630).[24] They do not: once we are careful to distinguish between the possibilities *sanctioned by quantum mechanics* and those not so sanctioned, we recognize that many-worlds interpretations give us no ground to deny non-actual possibilities—worlds, perhaps, in which the Schrödinger equation does not govern the dynamical evolution of physical systems, or worlds in which there are no physical systems at all. So while *quantum-mechanical* possibilities are indeed rendered actual in many-worlds interpretations, one would think that many other possible worlds remain non-actual. Based on what follows—endashed—right after the words cited earlier in this paragraph, I am confident that Putnam would not object to my petty remark here. But I nevertheless take this opportunity to remind the reader of this commonly overlooked fact.

In sum, these considerations move Putnam to conclude as follows:

> "What we are left with, if what I have said so far is right, is a conclusion that I initially found very distressing: *either GRW or some successor, or else Bohm or some successor, is the correct interpretation*—or... we will just fail to find a scientific realist interpretation which is acceptable." (631; emphasis in original)

Modulo my more hopeful assessment of many-worlds interpretations above, why would one take such a conclusion to be distressing? Putnam's disquiet arises from the fact that for both GRW and Bohmian mechanics, there exist at best some *ansätze* for how to turn these non-relativistic theories into fully Lorentz-invariant theories consistent with special relativity. At worst, as Putnam reaffirms (631) an earlier argument by Maudlin, none of these theories can ever be made *fully* Lorentz-invariant in that they rely, necessarily, on an 'absolute' time. If that were the case—that the only defensible solutions to the measurement problem must, by necessity, violate Lorentz-invariance and thus contradict relativity—, then that would indeed be distressing. For how could we then ever hope to achieve a unification in fundamental physics, i.e., to reconcile quantum physics with relativity theory, if the two families of theories stood some irreconcilable distance apart? Naturally, philosophers of science in the 'disunitist' camp, such as Cartwright, would hardly be fazed at the prospect of a fundamentally "dappled" world. But this is not the place to contribute to the debate between 'unitists' and 'disunitists', and so I will simply here grant that the coherence of fundamental physics demands a relaxed relationship between quantum physics and relativity theory. Incidentally, as a 'unitist', I believe that the reconciliation between the two is absolutely necessary and arguably the biggest outstanding challenge in fundamental physics. So I feel Putnam's distress.

[23] For some recent skirmishes in this ongoing debate, cf. Saunders et al. 2010.
[24] This suggestion perhaps goes back to the subtitle of DeWitt 1970.

Putnam's own—as he admits, speculative—resolution of the tension trades on a particular understanding of quantum cosmology, which, in his view, already contains an absolute time (632). His idea seems to be something like the following. First, in quantum cosmology, it is full spacetimes that have quantum states, and these quantum states "represent *different geometries of space-time*" (ibid.; emphasis in original). Presumably, this means that the relevant Hilbert space possesses a basis whose elements can be interpreted as eigenstates of operators that capture classical geometrical properties such as areas and volumes.[25] Putnam reassures us that

> [i]t might be that, before we 'superimpose', *each space-time is perfectly Einsteinian*—each space-time is a Minkowski space-time which knows nothing about any 'simultaneity'. (ibid.; emphasis in original)

So suppose it were indeed the case that each eigenstate, in an effort to appease the conflict between the quantum and relativity, would contain absolutely no structure that would permit the introduction of some absolute time. So where could absolute time come from? Since quantum cosmology is a *quantum* theory, and hence the superposition principle applies, the permissible states of the physical system will generally be some superposition of the 'geometrical' eigenstates. This is where Putnam's most speculative leap occurs:

> In effect, one superimposes whole space-times. And this superposition of space-times evolves in the background time… [This] would mean that, although Einstein would have to admit that there is such a thing as simultaneity, it comes from 'outside' any one well-defined space-time, it comes from the quantum mechanical 'interference' between whole space-times. (ibid.)

[25] For a discussion of what that could mean, cf. Wüthrich 2014.

Putnam seems to operate on an understanding of what quantum superpositions are that presupposes them to be situated at some specific 'external time' in which they then evolve. Somehow, if we are no longer confined to the well-behaved geometric eigenstates, we must accept that there exists a time external to spacetime and that quantum spacetimes can evolve in them and thus be in different states at different (external) times. Apart from the fact that it conjures up all those 'one-second-per-second' objections against multiple layers of time, I honestly cannot make sense of the picture that Putnam sketches here. The physical time in which any measureable physical processes transpire is the one frozen up in relativistic spacetimes. This strongly suggests that superpositions of quantum spacetimes require a rather different gloss from the one we standardly gave in non-relativistic quantum mechanics. I submit that the gloss with which we don superpositions in quantum cosmology and quantum gravity must be entirely ridden free of spatiotemporal undertones.[26] If I am right, then the tension between solutions to the measurement problem and relativistic demands on a fundamental theory would leave us in even greater distress than Putnam admits.

## 4 In conclusion

In this essay, I have defended the following four theses. First, against Cartwright, I have argued that Putnam's 1962 work on theoretical terms is compatible with his claim that quantum mechanics needs an interpretation. This need arises from my second—and altogether unoriginal—thesis that there *is* a measurement problem in quantum mechanics. Third, I largely agree with Putnam's own conclusions of 2005, except that I judge Everettian many-worlds theories to be more promising than he does and that I think that the apparent need of GRW and Bohmian mechanics of an absolute background time to be more unsettling than he does.

Fourth, as seems befitting for an essay in a collection celebrating his work, let me close by emphasizing just how instrumental Putnam was in convincing both philosophers and physicists that there is something rotten in the state of Copenhagen, and of quantum mechanics more widely. By the 1960s, most of the people working in fundamental physics had turned away from the foundational problems of quantum mechanics in favor of articulating the emerging standard model of particle physics—work that yielded more immediate results and benefits. Only a few intrepid souls such as John S Bell, Abner Shimony, and Hilary Putnam stubbornly continued to puzzle over these 'philosophical' problems at the heart of non-relativistic quantum mechanics that appeared increasingly quaint to so many others. Today, particularly as our attempts to incorporate gravity into a unified physical theory have so far led to few breakthroughs, most philosophers and many physicists recognize that there remains a central foundational problem that requires our attention. Putnam played an important role in getting us from there to here.

Let me illustrate this last point by retelling an episode Putnam relates. Here it is:

---

[26] For a discussion, cf., once again, Wüthrich 2014. In Wüthrich 2010, I argue that other attempts to reintroduce absolute time in quantum gravity fail.

In 1962 I had a series of conversations with a world-famous physicist (whom I will not identify by name). At the beginning, he insisted, 'You philosophers just *think* there is a problem with understanding quantum mechanics. We physicists have known better from Bohr on.'[suppressed footnote] After I forget how many discussions, we were sitting in a bar in Cambridge [(MA)], and he said to me, 'You're right. You've convinced me there is a problem here; it's a shame I can't take three months off and solve it.'

Fourteen years later [(1976)], the same physicist and I were together at a conference for a few days, and he opened his lecture at that conference (a lecture which explained to a general audience the exciting new theories of quarks) by saying, 'There is no Copenhagen interpretation of quantum mechanics. Bohr brainwashed a generation of physicists.' Evidently, he had undergone a considerable change of outlook. (2005, 619; emphasis in original)

I shall be less considerate than Putnam and frankly name the mystery physicist. As anyone with a web browser at their fingertip can find out, the mentioned conference was the 1976 Nobel Conference at Gustavus Adolphus College in Saint Peter, Minnesota, on "the nature of the physical universe", held in honor of Murray Gell-Mann, who had in 1969 won a Nobel Prize "for his contributions and discoveries concerning the classification of elementary particles and their interactions".[27] Why I am so confident that I correctly identified the conference and the physicist, you ask? Well, let me quote Gell-Mann from the proceedings of that conference:

… Niels Bohr brainwashed the whole generation of theorists into thinking that the job was done 50 years ago. (1979, 29)

Questions of identity aside, the story exemplifies how decisive a factor Putnam was in changing people's mind about the status of the measurement problem in the foundations of quantum mechanics, thereby illustrating my fourth thesis.

**Acknowledgments**

I am grateful to Hilary Putnam and to the editors of this volume for their patience. I would also like to thank Craig Callender and Nancy Cartwright for comments on an earlier draft.

[27] http://www.nobelprize.org/nobel_prizes/physics/laureates/1969/. For a list of speakers at the 1976 Gustavus Adolphus Nobel Conference, cf. https://gustavus.edu/events/nobelconference/archive/. Steven Weinberg, who also gave a talk at that conference, did not win his Nobel until 1979.